\documentclass[universe,review,accept,pdftex,moreauthors]{Definitions/mdpi} 
\firstpage{1} 
\pubvolume{11}
\issuenum{11}
\articlenumber{356}
\pubyear{2025}
\copyrightyear{2025}
\externaleditor{Øyvind Grøn} 
\datereceived{5 September 2025} 
\daterevised{14 October 2025} 
\dateaccepted{22 October 2025} 
\datepublished{28 October 2025} 
\hreflink{https://doi.org/10.3390/\\universe11110356} 

\usepackage{tikz}

\Title{Modified Gravity with Nonminimal Curvature--Matter Couplings: A Framework for Gravitationally Induced Particle~Creation}
\TitleCitation{Modified Gravity with Nonminimal Curvature--Matter Couplings: A Framework for Gravitationally Induced Particle Creation}

\Author{{Francisco} {S. N.} Lobo $^{1,2,}$*, Tiberiu Harko $^{3,4}$ and Miguel {A. S.} Pinto $^{1,2}$}

\AuthorNames{Francisco S. N. Lobo, Tiberiu Harko and Miguel A. S. Pinto}

 \AuthorCitation{Lobo, F.S.N.; Harko, T.; Pinto, M.A.S.}
\address{%
$^{1}$ \quad Instituto de Astrof\'{i}sica e Ci\^{e}ncias do Espa\c{c}o, Faculdade de Ci\^encias da Universidade de Lisboa, Edif\'{i}cio C8, Campo Grande, {PT}-1749-016 Lisbon, Portugal; {mapinto@fc.ul.pt}\\
$^{2}$ \quad Departamento de F\'{i}sica, Faculdade de Ci\^{e}ncias, Universidade de Lisboa, Edifício C8, Campo Grande, {PT}{-}1749-016 Lisbon, Portugal\\
$^{3}$ \quad Faculty of Physics, Babe\c s-Bolyai University, 1 Kog\u alniceanu Street, 400084 Cluj-Napoca, Romania; {tiberiu.harko@aira.astro.ro}\\
$^{4}$ \quad Astronomical Observatory, 19 Cire\c silor Street, 400487 Cluj-Napoca, Romania}

\corres{Correspondence: fslobo@fc.ul.pt}

\abstract{Modified gravity theories with a nonminimal coupling between curvature and matter offer a compelling alternative to dark energy and dark matter by introducing an explicit interaction between matter and curvature invariants. Two of the main consequences of such an interaction are the emergence of an additional force and the non-conservation of the energy--momentum tensor, which can be interpreted as an energy exchange between matter and geometry. By adopting this interpretation, one can then take advantage of many different approaches in order to investigate the phenomenon of gravitationally induced particle creation. One of these approaches relies on the so-called irreversible thermodynamics of open systems formalism. By considering the scalar--tensor formulation of one of these theories, we derive the corresponding particle creation rate, creation pressure, and entropy production, demonstrating that irreversible particle creation can drive a late-time de Sitter acceleration through a negative creation pressure, providing a natural alternative to the cosmological constant. Furthermore, we demonstrate that the generalized second law of thermodynamics holds: the total entropy, from both the apparent horizon and enclosed matter, increases monotonically and saturates in the de Sitter phase, imposing constraints on the allowed particle production dynamics. Furthermore, we present brief reviews of other theoretical descriptions of matter creation processes. Specifically, we consider approaches based on the Boltzmann equation and quantum-based aspects and discuss the generalization of the Klein--Gordon equation, as well as the problem of its quantization in time-varying gravitational fields. Hence, gravitational theories with nonminimal curvature--matter couplings present a unified and testable framework, connecting high-energy gravitational physics with cosmological evolution and, possibly, quantum gravity, while remaining consistent with local tests through suitable coupling functions and screening mechanisms.}
\keyword{modified gravity; nonminimal curvature--matter couplings; particle creation; scalar--tensor representation; irreversible thermodynamics of open systems; Boltzmann equation; quantum field theory in curved spacetimes}

\begin{document}




\section{Introduction}

The discovery of the late-time accelerated expansion of the universe has profoundly challenged our understanding of fundamental physics. This unexpected phenomenon, first observed through Type Ia supernovae data {\cite{Riess:1998cb,Perlmutter:1998np}} and later corroborated by precise measurements of the cosmic microwave background (CMB), baryon acoustic oscillations (BAO), and large-scale structure surveys {\cite{Planck:2018vyg,Eisenstein:2005su}}, has sparked an intense debate: Is General Relativity (GR), our well-established theory of gravitation, still valid on cosmological scales? Or does this evidence hint at the necessity for new gravitational physics?
These fundamental questions have motivated a growing body of work exploring whether the Einstein field equations require modification in the infrared (IR) regime. The standard cosmological model, $\Lambda$CDM, while phenomenologically successful, relies on the existence of two mysterious components—dark energy and dark matter—neither of which has been directly detected in laboratory experiments. As a result, theorists have suggested that the cosmic acceleration may not be due to an exotic matter-energy content but rather signals a breakdown of GR at large scales \cite{Clifton:2011jh,Nojiri:2010wj}.

In fact, cosmology offers a unique setting to test gravity far beyond the regimes where GR has been extensively confirmed, such as in the solar system and binary pulsars {\cite{Will:2014kxa}}. On cosmological scales, discrepancies between theoretical predictions and observational data provide fertile ground to probe deviations from Einstein gravity \cite{Lobo:2008sg,Avelino:2016lpj,CANTATA:2021asi,CosmoVerse:2025txj}. One promising direction is to consider generalizations of the Einstein--Hilbert action by introducing higher-order curvature invariants. These include terms such as $R^2$, $R_{\mu \nu}R^{\mu \nu}$, $R_{\alpha \beta \mu \nu}R^{\alpha \beta \mu \nu}$ (where $R$, $R_{\mu\nu}$, and $R_{\alpha\beta\mu\nu}$ are the Ricci scalar, Ricci tensor, and Riemann tensor, respectively), the topological Gauss--Bonnet term $\varepsilon^{\alpha\beta\mu\nu} R_{\alpha\beta\gamma\delta} R^{\gamma\delta}{}_{\mu\nu}$, and the self-contraction of the Weyl tensor $C_{\alpha \beta \mu \nu} C^{\alpha \beta \mu \nu}$. These corrections naturally arise in the low-energy effective action of many different approaches to quantum gravity, such as string theory or loop quantum gravity, and serve as viable candidates for ultraviolet (UV) or IR completions of GR {\cite{Stelle:1976gc,Birrell:1982ix}}.
Indeed, there are multiple physical motivations for considering such extensions. Firstly, higher-curvature terms can regularize singularities found in classical solutions of GR, providing a more realistic depiction of spacetime near high-curvature regions like black hole cores or the Big Bang ~\cite{Sotiriou:2008rp}. Secondly, these modifications may represent the leading-order quantum corrections to classical gravity and thus bridge the gap between GR and a fundamental quantum theory of spacetime~\cite{DeFelice:2010aj}.

\textls[-15]{Among the simplest and most studied modifications of GR is $f(R)$ gravity, formulated by Hans Buchdahl in the 1970s \cite{Buchdahl:1970ynr}. In this framework, the standard Einstein--Hilbert action is generalized by replacing the Ricci scalar $R$ with an arbitrary function $f(R)$, yielding \mbox{the following {action:}}}
\begin{equation}
	S = \int d^4x\, \sqrt{-g} \left[ \frac{f(R)}{2\kappa^2} + \mathcal{L}_m(g_{\mu\nu}, \psi) \right],
\end{equation}
where $\kappa^2 = 8\pi G$, with $G$ being the universal gravitational constant, $g$ is the determinant of the metric tensor $g_{\mu\nu}$, and $\mathcal{L}_m$ represents the matter Lagrangian, which generally depends on the metric and on a collection of matter fields and their derivatives $\psi$. This class of theories retains general covariance and introduces new dynamics through the functional dependence on $R$. The appeal of $f(R)$ gravity lies in its ability to encompass a wide range of gravitational phenomena with relatively minimal departures from the geometric structure of GR~\cite{Sotiriou:2008rp,DeFelice:2010aj}.
\textcolor{black}{Now, varying the action with respect to the metric yields
	$
	F(R)R_{\mu\nu} - \tfrac{1}{2}f(R)g_{\mu\nu}
	- \nabla_\mu\nabla_\nu F(R) + g_{\mu\nu}\Box F(R) = \kappa T_{\mu\nu}$, where $F \equiv df/dR$.
	These are \emph{fourth-order} differential equations in the metric $g_{\mu\nu}$. The reason is that $R$ itself contains second derivatives of the metric,
	$R \sim \partial^2 g$, while $F(R)$ depends on $R$. The terms $\nabla_\mu\nabla_\nu F(R)$ and $\Box F(R)$ therefore involve derivatives of $F(R)$, producing terms of the form
	$
	\nabla\nabla F(R) \sim \nabla\nabla(\partial^2 g) \sim \partial^4 g.
	$
	Hence, the dependence of the Lagrangian on a nonlinear function of $R$ introduces two additional derivatives of the metric compared with GR, where $f(R)=R$ and the corresponding equations remain second order.}
Taking the trace of these equations yields a dynamical equation for an extra scalar degree of freedom: $3 \Box F + F R - 2f = \kappa^2 T$, where $T$ is the trace of the energy--momentum tensor, and $\Box$ is the d'Alembertian operator. This equation shows that the Ricci scalar is no longer an algebraic function of the matter sources, as in GR, but instead evolves dynamically. In fact, $f(R)$ theories can be recast as scalar--tensor theories, with the scalar field $\phi \sim F$ additionally mediating the gravitational interaction~\cite{Sotiriou:2008rp,Sotiriou:2006qn}.

The introduction of this new scalar degree of freedom in $f(R)$ gravity allows the theory to naturally explain the late-time cosmic acceleration without the need to explicitly include a cosmological constant or a dark energy component~\cite{Capozziello:2002rd,Carroll:2003wy,Nojiri:2003ft}. In this framework, the scalar field effectively behaves as a dynamical dark energy component, modifying the expansion history of the universe and providing a mechanism for accelerated expansion that evolves over time. However, these modifications to gravity are not without challenges: generally, light scalar fields mediate long-range fifth forces that can produce measurable deviations from GR. Such deviations are tightly constrained by precision tests within the solar system and experiments probing the equivalence principle~\cite{Will:2014kxa}. Consequently, simple or naive implementations of $f(R)$ gravity risk conflicting with these local gravity observations.
To resolve this tension, screening mechanisms have been proposed, among which is the chameleon mechanism~\cite{Khoury:2003rn,Khoury:2003aq}, where the effective mass of the scalar field dynamically depends on the local matter density: in high-density environments, such as within the solar system, the scalar field acquires a large effective mass, thereby suppressing the range of the fifth force and effectively restoring standard GR predictions. On the other hand, in the low-density cosmological background, the scalar field remains light and can exert significant influence on the cosmic expansion, thus driving the observed acceleration. This environment-dependent behavior enables $f(R)$ gravity models to satisfy stringent local tests while still addressing cosmological phenomena.

There exist several formulations of $f(R)$ gravity, which differ primarily in how 
the affine connection is treated. In the \emph{metric formalism}, the connection 
is assumed to be the Levi-Civita connection associated with the metric, and the 
action variation is performed solely with respect to the metric tensor. This 
approach leads to fourth-order differential field equations in the metric 
components, as mentioned above. Alternatively, the \emph{Palatini formalism} 
treats the metric and the connection as independent variables; the variation is performed with respect to both, yielding second-order field equations that differ from those obtained in the metric formalism~\cite{Olmo:2011uz}. The \emph{metric-affine formalism} further generalizes the theory by allowing the matter action to depend explicitly on the independent connection, thereby introducing additional geometric degrees of freedom ~\cite{Sotiriou:2006qn}. More recently, a \emph{hybrid metric-Palatini formalism} has been proposed, which combines elements of both the metric and Palatini approaches ~\cite{Harko:2011nh,Capozziello:2015lza,Harko:2020ibn}. This hybrid theory interpolates between these two extremes, providing novel cosmological dynamics and late-time acceleration while remaining consistent with local gravity constraints \cite{Capozziello:2012ny,Capozziello:2012qt,Capozziello:2013uya}.

While $f(R)$ gravity represents a significant departure from Einstein’s theory by modifying the geometric sector of the action, it remains minimal in its interaction with matter. However, there exists a natural and compelling extension of this framework that introduces a direct coupling between curvature and matter. This class of theories, known as curvature--matter coupling models, infers that matter may interact nonminimally with curvature invariants, such as the Ricci scalar $R$, leading to further deviations from standard dynamics.
One such extension was proposed in Refs.~\cite{Gonner:1976gq,Gonner:1984zx,Bertolami:2007gv}, where $f(R)$ gravity was generalized by adding a nonminimal linear coupling between the Ricci scalar and the matter Lagrangian. The action in this model is given by the following:
\begin{equation}
	S = \int d^4x\, \sqrt{-g} \left\{ \frac{1}{2} f_1(R) + \left[1 + \lambda f_2(R)\right] \mathcal{L}_m \right\},
	\label{action2}
\end{equation}
where $f_1(R)$ and $f_2(R)$ are arbitrary functions of the Ricci scalar and $\lambda$ is a coupling constant quantifying the strength of the curvature--matter interaction. In the limit $\lambda \to 0$, standard $f(R)$ gravity is recovered, while for general $\lambda$, the matter sector dynamically responds not only to the metric but also to the curvature scalar directly.

This type of coupling leads to profound consequences for both theoretical consistency and phenomenology \cite{Harko:2024sea}. Notably, the energy--momentum tensor of matter is no longer conserved; instead, its divergence is proportional to the gradient of $R$, implying that particles follow non-geodesic paths~\cite{Bertolami:2007gv, Harko:2008qz}. This feature provides an intriguing avenue to model dark matter and dark energy effects without invoking exotic matter components. For instance, the additional force emerging from the curvature--matter coupling can mimic the dark matter behavior in galactic rotation curves~\cite{Bertolami:2007vu} and contribute to the late-time cosmic acceleration~\cite{Bertolami:2007gv, Harko:2010mv}. In addition, it has been recently shown that such a modified gravity theory alleviates the Hubble tension \cite{BV1} and is in agreement with some of the latest cosmological data \cite{BV2}. Gravitational wave polarizations \cite{BV3}, matter density \mbox{perturbations \cite{BV4}}, and inflationary scenarios \cite{BV5} were also explored within this theoretical framework.
Furthermore, these models open the possibility of distinguishing different gravitational regimes by the environment or the nature of the matter content, enriching the landscape of viable modified gravity theories. They offer an appealing framework to explore the unification of gravitational and matter dynamics at a more fundamental level, while being subject to stringent constraints from local physics and cosmological observations.
The study of nonminimal curvature--matter coupling theories thus marks an important step in the broader program of testing the foundations of gravitational theory and extending our understanding of the gravitational interaction beyond \mbox{the Einsteinian paradigm.}

\textls[-15]{An especially intriguing aspect of theories with nonminimal curvature--matter couplings is their ability to incorporate particle production processes within a cosmological setting. Rather than invoking external mechanisms, these models offer a geometrically motivated interpretation of matter creation, grounded in the thermodynamic treatment of open systems. In particular, the formalism pioneered by Prigogine and collaborators~\cite{Prigogine:1988jax,Prigogine:1989zz} interprets particle production as an irreversible process, characterized by entropy growth and an effective pressure resembling bulk viscosity. Within the context of nonminimal curvature--matter couplings, the non-conservation of the energy--momentum tensor naturally reflects an ongoing exchange of energy between geometry and matter—an effect that closely parallels the dynamics of particle creation~\cite{Harko:2014pqa,Harko:2015pma,Pinto:2022tlu,Cipriano:2023yhv}. This synthesis of modified gravity and irreversible thermodynamics provides a coherent framework to model cosmological matter generation and offers promising avenues for confronting theoretical predictions with current and future observational data. This will be explored in detail throughout \mbox{this work. }}

This work is organized in the following manner: In Section \ref{secII}, we present the modified gravitational action with a specific nonminimal curvature--matter coupling and derive the corresponding field equations; furthermore, we discuss the emergence of an extra force due to the non-conservation of the energy--momentum tensor; then, we reformulate the theory in terms of a dynamically equivalent scalar--tensor theory and discuss the potential applications in cosmology and astrophysics.
In Section \ref{secIII}, we explore in depth the formalism of irreversible thermodynamics of open systems applied to cosmology as a means to describe irreversible particle creation, emphasizing its role in the dynamics of the cosmic expansion. We also discuss the total entropy of the universe, develop a covariant thermodynamic description, including the entropy flux and its divergence in the presence of nonminimal curvature--matter coupling, explore the evolution of temperature, and provide an interpretation of the second law's validity and its constraints on particle production rates. Moreover, in order to ensure a more complete review, we further examine alternative approaches to gravitationally induced particle creation, with particular attention to Boltzmann and quantum-based frameworks, explored in Sections \ref{secIV} and \ref{secV}, respectively. Finally, we summarize and discuss our results and conclude in Section \ref{secconclusion}.

\section{Modified Gravity with a Nonminimal Curvature--Matter Coupling: Theory and Implications}
\label{secII}

\subsection{Action and Field Equations}

The action for the nonminimal curvature--matter coupling theory provided above, but rewritten here for self-completeness and self-consistency, is given by the following \cite{Bertolami:2007gv}:
\begin{equation}
	S=\int \left\{\frac{1}{2}f_1(R)+\left[1+\lambda f_2(R)\right]{\cal L}_{m}\right\} \sqrt{-g}\;d^{4}x~,
\end{equation}
where \( f_1(R) \) and \( f_2(R) \) are arbitrary functions of the Ricci scalar \( R \), \( \lambda \) is a coupling constant, and \( {\cal L}_m \) is the matter Lagrangian density. Varying this action with respect to the metric yields a set of modified field equations, written as follows:
%
\begin{equation}
\label{FE}
\begin{array}{rr}
	F_1(R)R_{\mu \nu }-\frac{1}{2}f_1(R)g_{\mu \nu }-\nabla_\mu
	\nabla_\nu \,F_1(R)+g_{\mu\nu}\square F_1(R)
	=-2\lambda F_2(R){\cal L}_m R_{\mu\nu}		\\
	{~}\\
	+2\lambda(\nabla_\mu
	\nabla_\nu-g_{\mu\nu}\square){\cal L}_m F_2(R)
	+[1+\lambda f_2(R)]T_{\mu \nu }^{(m)}~,
\end{array}
\tag{4}
\end{equation}
where \( F_i(R) = df_i/dR \), while the matter energy--momentum tensor $T_{\mu \nu}^{(m)}$ is defined in the standard way
\begin{equation}
	T_{\mu \nu
	}^{(m)}=-\frac{2}{\sqrt{-g}}\frac{\delta(\sqrt{-g}\,{\cal L}_m)}{\delta g^{\mu\nu}}.
	\tag{5}
\end{equation}

However, due to the explicit nonminimal curvature--matter coupling, the standard conservation law \( \nabla^\mu T_{\mu\nu}^{(m)} = 0 \) is no longer valid. Taking the covariant divergence of the field equations, applying the Bianchi identities \( \nabla^\mu \left[ R_{\mu\nu} -(1/2)g_{\mu\nu}R\right] = 0 \), and using the identity \( (\Box \nabla_\nu - \nabla_\nu \Box) F_i = R_{\mu\nu} \nabla^\mu F_i \), one arrives at the modified conservation law, written as follows:
\begin{equation}
	\nabla^\mu T_{\mu \nu }^{(m)}=\frac{\lambda F_2}{1+\lambda
		f_2}\left[g_{\mu\nu}{\cal L}_m- T_{\mu \nu
	}^{(m)}\right]\nabla^\mu R ~.
	\tag{6}
\end{equation}

This indicates that an exchange of energy and momentum can occur between the matter content and the geometrical degrees of freedom \cite{Bertolami:2007gv,Harko:2010mv}. 

Interestingly, similar coupling structures emerge in scalar--tensor theories when conformal transformations are applied to move between the Jordan and Einstein frames~\cite{Faraoni:2004pi}, and they are also common in low-energy effective actions derived from string theory~\cite{Maeda:1988ab}. In the limit \( \lambda \to 0 \), the standard minimal coupling between curvature and matter is recovered, and the energy--momentum conservation follows from the diffeomorphism invariance of the matter sector and from such a minimal coupling.

Various formulations of this theory have been studied, including both metric and Palatini approaches. In particular, the Palatini version of the nonminimal curvature--matter coupling theory presented here has been explored in~\cite{Harko:2010hw}, where the connection and metric are treated as independent variables, leading to a distinct set of field equations with interesting phenomenological consequences.

\subsection{Equations of Motion and the Extra Force}

The non-conservation of the energy--momentum tensor manifests physically as a deviation from geodesic motion. For a perfect fluid, the equation of motion for a fluid element reads as follows:
\begin{equation}\label{eqmot}
	\frac{Du^{\alpha }}{ds} \equiv \frac{du^{\alpha }}{ds}+\Gamma _{\mu
		\nu }^{\alpha }u^{\mu }u^{\nu }=f^{\alpha },
		\tag{7}
\end{equation}
where the extra force \( f^\alpha \), orthogonal to the four-velocity, is given by
\begin{equation}\label{extra}
	f^{\alpha } = \frac{1}{\rho +p}\Bigg[\frac{\lambda
		F_2}{1+\lambda f_2}\left({\cal L}_m-p\right)\nabla_\nu
	R+\nabla_\nu p \Bigg] h^{\alpha \nu },
	\tag{8}
\end{equation}
with \( h_{\mu \nu }=g_{\mu \nu }+u_{\mu }u_{\nu } \) being the projection tensor. This extra force introduces significant phenomenological consequences. Interestingly, the form of this force depends on the choice of the matter Lagrangian: whether one uses \( {\cal L}_m = -\rho \) or \( {\cal L}_m = p \) leads to qualitatively different outcomes \cite{Bertolami:2008ab}.

\subsection{Scalar--Tensor Representation}

An elegant and insightful reformulation of this nonminimally coupled theory can be achieved by recasting it in a scalar--tensor framework. Introducing the scalar fields \( \phi \) and \( \psi(\phi) \equiv f_1^{\prime}(\phi) \), the action can be written as follows (\textcolor{black}{we refer the reader to Ref. \cite{Sotiriou:2008it} for \mbox{the specific details}}):
\vspace{-6pt}
\begin{equation}
	S=\int d^4x \sqrt{-g} \left[ \frac{\psi R }{2} -V(\psi)\, +U(\psi) \mathcal{L}_m \right],
	\tag{9}
\end{equation}
where the scalar potentials \( V(\psi) \) and \( U(\psi) \) are defined as
\begin{equation}
	V(\psi) = \frac{\phi(\psi) f_1^{\prime }\left[ \phi (\psi ) \right] -f_1\left[ \phi( \psi ) \right] }{2},
	\qquad
	U( \psi) = 1+\lambda f_2\left[ \phi( \psi ) \right].
	\tag{10}
\end{equation}

\textls[-15]{The original action and this scalar--tensor form are dynamically equivalent provided that \( f_1^{\prime \prime}(R) \neq 0 \) \cite{Bertolami:2007gv,Harko:2010mv}. This representation facilitates comparisons with known scalar--tensor models, provides useful tools for studying cosmological dynamics, and clarifies the interpretation of the new degrees of freedom as effective scalar fields interacting with matter.}

In the scalar--tensor representation of the theory, the divergence of the energy--momentum tensor is modified due to the coupling between the scalar field \( \psi \) and the matter Lagrangian. Specifically, one finds the following:
\begin{equation}
\nabla_\mu T{^\mu}{_{\nu}}=\left(\delta_\nu^\mu\mathcal{L}_\text{m}-T{^\mu}{_{\nu}}\right) \nabla_\mu \ln U(\psi),
\tag{11}
\end{equation}

This equation indicates the exchange of energy--momentum between matter and the scalar field \( \psi \), mediated by the self-interaction potential \( U(\psi) \) \cite{Harko:2010mv}.


As before, assuming that the matter sector is described by a perfect fluid with energy density \( \rho \), pressure \( p \), and four-velocity \( u^\mu \), the energy balance equation becomes the~following:
\begin{equation}
	\dot{\rho} + 3H(\rho + p) + \left(\rho-\mathcal{L}_m\right) \frac{d}{dt} \ln U(\psi) = 0 ~,
	\tag{12}
\end{equation}
{where} \( \dot{} = u^\mu \nabla_\mu \) and $\nabla_{\mu} u^{\mu}=3H$, with \( H \) being the Hubble parameter in a Friedmann--Robertson--Walker (FRW) background.

\subsection{Applications and Outlook}

The implications of nonminimal couplings between curvature and matter are broad and diverse. In cosmology, these models have been employed to explain the late-time accelerated expansion of the universe without the need for a cosmological constant~\cite{Bertolami:2007gv,Harko:2010mv,Harko:2011kv}. From the perspective of gravitational thermodynamics, the coupling induces a non-equilibrium framework characterized by entropy production~\cite{,Harko:2014pqa,Harko:2015pma,Harko:2014gwa}. Astrophysically, these modifications affect the internal structure of stellar objects, provide alternative explanations for galactic rotation curves without invoking dark matter~\cite{Harko:2018ayt}, and influence the properties and stability of compact configurations including neutron stars \cite{Lobato:2020fxt,Pretel:2021kgl,Lobato:2021ehf} and wormholes~\cite{MontelongoGarcia:2010xd,Garcia:2010xb}. Consequently, nonminimal curvature--matter coupling theories offer a versatile theoretical framework for investigating modified gravity scenarios with rich phenomenological consequences spanning cosmology and astrophysics.


\section{Irreversible Thermodynamics of Open Systems and Particle Creation}\label{secIII}

In modified theories of gravity featuring a nonminimal curvature--matter coupling, the interaction between geometry and matter can be interpreted through the formalism of irreversible thermodynamics of open systems. In this framework, matter creation is understood as an irreversible process occurring on cosmological scales, intrinsically related to the non-conservation of the matter energy--momentum tensor~\cite{Prigogine:1988jax,Prigogine:1989zz,Harko:2014pqa}. Unlike the standard adiabatic conservation equation in GR, the modified energy conservation law acquires an additional source term. This term represents an irreversible particle creation rate, modifying the conventional balance equations that govern the evolution of cosmic fluids~\cite{Lima:1992np}. According to irreversible thermodynamics, particle creation acts as an entropy source that generates an entropy flux, thereby altering the temperature evolution of the cosmological fluid from its standard adiabatic behavior \cite{Calvao:1991wg}. This unified framework merges geometric modifications of gravity with the thermodynamics of open systems, yielding a consistent phenomenological approach to cosmological particle creation that can be tested against observational data.

In what follows, we will consider this theoretical framework in order to explore gravitationally induced particle production in the nonminimal curvature--matter coupling theory in its scalar--tensor representation we presented before.

\subsection{Thermodynamic Framework in an Expanding Universe}

The formal inclusion of irreversible matter creation processes within GR was pioneered by Prigogine and collaborators \cite{Prigogine:1988jax} and later developed in a broader cosmological context. The central idea is that the standard conservation of energy--momentum in Einstein’s equations is modified due to a continuous, irreversible creation of particles, interpreted as an energy flow from the gravitational field to the matter sector.

Consider a spatially flat Friedmann--Lemaître--Robertson--Walker (FLRW) spacetime, written as follows:
\begin{equation}
	ds^2 = dt^2 - a^2(t)(dx^2 + dy^2 + dz^2),
	\tag{13}
\end{equation}
where \( a(t) \) is the scale factor. The universe is filled with a perfect fluid characterized by a particle number \( N \), energy density \( \rho \), pressure \( p \), and volume \( V \propto a^3 \). The total particle number density is \( n = N/V \). The first law of thermodynamics for this open cosmological system is generalized as follows \cite{Prigogine:1988jax}:
\begin{equation}
	\frac{d}{dt}(\rho a^3) + p \frac{d}{dt}a^3 = \frac{dQ}{dt} + \frac{\rho + p}{n} \frac{d}{dt}(n a^3),
	\tag{14}
\end{equation}
where \( dQ \) is the heat exchanged with the environment. In the adiabatic limit \( dQ = 0 \), this leads to the following:
\begin{equation}
	\dot{\rho} + 3H(\rho + p) = \frac{\rho + p}{n} \left( \dot{n} + 3Hn \right), \label{eq:energy_balance}
	\tag{15}
\end{equation}
where \( H = \dot{a}/a \) is the Hubble parameter. Introducing the particle creation rate \( \Gamma \) via \cite{Harko:2014pqa}.
\begin{equation}
	\dot{n} + 3Hn = \Gamma n,
    \label{n}
    \tag{16}
\end{equation}

Equation~\eqref{eq:energy_balance} can be recast into a modified energy conservation equation, written \linebreak as follows:
\begin{equation}
	\dot{\rho} + 3H(\rho + p) = (\rho + p)\Gamma. \label{eq:modified_energy}
	\tag{17}
\end{equation}

The explicit form of \(\Gamma\) depends on the gravity theory one considers. For the nonminimal curvature--matter coupling theory presented in Section \ref{secII}, $\Gamma$ depends on the curvature--matter coupling function \(U(\psi)\) and is given by the following:
\begin{equation}
		\Gamma = \frac{\mathcal{L}_m-\rho}{\rho + p} \frac{d}{dt} \ln U(\psi).
		\tag{18}
	\end{equation}

The first law of thermodynamics for an open system with adiabatic particle creation can also be equivalently expressed as follows:
	\begin{equation}
		\frac{d}{dt} (\rho a^3) + (p + p_c) \frac{d}{dt} a^3 = 0,
		\tag{19}
	\end{equation}
	where \(a\) is the scale factor and \(p_c\) denotes the creation pressure~\cite{Prigogine:1988jax,Lima:1992np}. This relation can be rewritten in differential form as follows:
	\begin{equation}
		\dot{\rho} + 3 H (\rho + p + p_c) = 0,
		\tag{20}
	\end{equation}
	where the creation pressure encodes the thermodynamic effect of particle production on the cosmic fluid \cite{Calvao:1991wg}. The creation pressure is related to the particle creation rate by\linebreak  the following:
	\begin{equation}
		p_c = - \frac{\rho + p}{3} \frac{\Gamma}{H}.
		\tag{21}
        \label{pc}
	\end{equation}
	
	Substituting the explicit expression for \(\Gamma\), the creation pressure in the presence of curvature--matter coupling becomes the following:
	\begin{equation}
		p_c = \frac{\rho-\mathcal{L}_m}{3H} \frac{d}{dt} \ln U(\psi).
		\tag{22}
	\end{equation}
	
	This result demonstrates how the coupling between geometry and matter generates an effective negative pressure through irreversible particle creation, which may have significant implications for the thermal history and accelerated expansion of the universe~\cite{Harko:2014pqa}.

\subsection{Entropy Evolution}
	
The fundamental principles of the thermodynamics of open systems establish that the total change in entropy \(dS\) can be decomposed into two distinct parts: the entropy flow term \(d_e S\) and the entropy creation term \(d_i S\). The entropy flow, \(d_e S\), corresponds to the exchange of entropy between the system and its surroundings, while the entropy creation term, \(d_i S\), arises from irreversible processes occurring within the system itself~\cite{Prigogine:1988jax}. Thus, the total entropy change is expressed as follows:
	\begin{equation}
		dS = d_e S + d_i S,
		\tag{23}
	\end{equation}
with the second law of thermodynamics imposing the constraint \(d_i S > 0\), which reflects the irreversible nature of entropy production.
	
For closed thermodynamic systems undergoing adiabatic transformations, the entropy change vanishes identically, implying \(dS = 0\) and \(d_i S = 0\). However, in the presence of curvature--matter coupling, which effectively induces matter creation, the system is no longer closed, and irreversible processes generate a positive entropy creation rate, leading to an overall increase in the total entropy of the cosmic fluid~\cite{Prigogine:1988jax,Prigogine:1989zz,Harko:2014pqa}. Moreover, in a homogeneous and isotropic universe, the entropy flow term \(d_e S\) is typically zero due to the absence of entropy exchange with an external environment, such that all entropy change arises solely from internal irreversible processes.
	
The irreversible entropy production associated with particle creation is quantitatively captured by the time rate of change of the entropy creation term, which can be expressed as follows~\cite{Prigogine:1988jax}:
	\begin{equation}  \label{entropy_production}
		T \frac{d_i S}{dt} = T \frac{dS}{dt} = \frac{h}{n} \frac{d}{dt} (n a^3) - \mu \frac{d}{dt} (n a^3) \geq 0,
		\tag{24}
	\end{equation}
where \(T\) is the temperature of the fluid, \(h = \rho + p\) is the enthalpy per unit volume, \(\mu\) denotes the chemical potential, \(n\) is the particle number density, and \(a\) is the cosmological scale factor. This inequality reflects the second law, demanding non-negative entropy production in the presence of particle creation processes.
	
Using the particle number balance equation introduced previously, the following\linebreak  is obtained:
	\[
	\dot{n} + 3 H n = \Gamma n,
	\]
the entropy production rate can be rewritten in a compact form as
	\begin{equation}
		\frac{dS}{dt} = \frac{S}{n} (\dot{n} + 3 H n) = \Gamma S \geq 0,
		\label{entropy_rate}
		\tag{25}
	\end{equation}
where \(S\) is the total entropy of the system. Integrating this equation yields an exponential growth of entropy due to particle creation,
	\begin{equation}
		S(t) = S_0 \, e^{\int_0^t \Gamma(t') dt'},
		\tag{26}
	\end{equation}
with \(S_0 = S(0)\) representing the initial entropy at the initial time.
	
Finally, employing the explicit form of the particle creation rate \(\Gamma\) from the scalar--tensor representation of the linear curvature--matter coupling model, the entropy evolution equation takes the following form:
\begin{equation}
		\frac{1}{S} \frac{dS}{dt} = \frac{\mathcal{L}_m-\rho}{\rho + p} \frac{d}{dt} \ln U(\psi),	\label{entropy_scalar_tensor}
		\tag{27}
	\end{equation}
explicitly connecting the entropy production rate to the dynamics of the curvature--matter coupling function \(U(\psi)\). This framework not only provides a thermodynamically consistent description of matter creation but also links microscopic entropy generation to cosmological dynamics and the evolution of the universe~\cite{Harko:2014pqa,Prigogine:1989zz,Prigogine:1988jax}.

\subsection{Entropy Flux Four-Vector and Irreversible Thermodynamics}

In the framework of relativistic thermodynamics and cosmology with particle creation, the entropy flux four-vector \( S^{\mu} \) plays a fundamental role in characterizing the irreversible processes associated with entropy generation in the universe. For a relativistic fluid with particle number density \( n \), four-velocity \( u^{\mu} \), and specific entropy \( \sigma = S/N \), the entropy flux is defined as follows:
\begin{equation}
	S^{\mu} = n \sigma u^{\mu},
	\tag{28}
\end{equation}
where \( S \) is the total entropy and \( N \) the total number of particles. This expression generalizes the concept of entropy in a covariant form compatible with the structure of GR~\cite{Eckart:1940zz,Weinberg:1972kfs}.

To satisfy the second law of thermodynamics, the divergence of the entropy flux vector must be non-negative, written as follows:
\begin{equation}
	\nabla_{\mu} S^{\mu} \geq 0,
	\tag{29}
\end{equation}
ensuring that entropy production is always positive or zero in accordance with thermodynamic principles.

The evaluation of \( \nabla_{\mu} S^{\mu} \) in the context of gravitational particle creation relies on thermodynamic identities. Employing the Gibbs relation and the definition of the chemical potential \( \mu \), one obtains the following:
\begin{equation}
	\nabla_{\mu} S^{\mu} = \frac{1}{T} \left( \dot{n} + 3 H n \right) \left( \frac{h}{n} - \mu \right),
	\tag{30}
\end{equation}
where \( T \) is the temperature, \( h = \rho + p \) is the enthalpy density of the fluid, \( \rho \) is the energy density, \( p \) the pressure, and \( H \) the Hubble parameter. The quantity \( \dot{n} + 3 H n \) accounts for the change in particle number due to expansion and creation.

In theories with curvature--matter coupling, the energy balance equation implies a modified continuity equation and leads to an explicit expression for the particle creation rate \( \Gamma \). Substituting the earlier derived expression for $\Gamma$, we obtain the entropy production rate in terms of the scalar field dynamics and coupling functions, written as follows:
\begin{equation}
	\nabla_{\mu} S^{\mu} = \frac{n\left(\mathcal{L}_m-\rho\right) }{T (\rho + p)} \left( \frac{h}{n} - \mu \right) \frac{d}{dt} \ln U(\psi).
	\tag{31}
\end{equation}

This result connects the geometric structure of spacetime, encoded in the scalar field \( \psi \) and the coupling function \( U(\psi) \), to the irreversible thermodynamic properties of the universe. It shows explicitly how the nonminimal interaction between matter and curvature acts as a source of entropy, providing a deeper insight into the interplay between gravity, thermodynamics, and the arrow of time in cosmology~\cite{Harko:2014pqa,Prigogine:1988jax,Eckart:1940zz}.

\subsection{Temperature Evolution in Nonminimal Curvature--Matter Coupled Cosmologies}

In the thermodynamic description of cosmological models with particle creation, a fundamental role is played by the temperature evolution of the matter fluid. A general thermodynamic system is typically described by two independent variables: the particle number density \( n \) and the temperature \( T \). If the system is assumed to be in thermodynamic equilibrium, all other quantities of interest, such as the energy density \( \rho \) and pressure \( p \), can be derived from the equations of state, written as follows:
\begin{equation}
	\rho = \rho(n, T), \qquad p = p(n, T). \label{eq:rho-p-eos}
	\tag{32}
\end{equation}

These functional relations reflect the equilibrium microphysics of the fluid. The evolution of the temperature is then directly tied to the dynamics of the number density and the expansion of the universe, as well as to the underlying gravitational interaction when matter creation is induced by geometry.

Starting from the generalized energy balance equation in the presence of a particle creation term, one can write the following:
\begin{equation}
	\frac{\partial \rho}{\partial n} \dot{n} + \frac{\partial \rho}{\partial T} \dot{T} + 3H(\rho + p) = \Gamma n, \label{eq:energy-balance}
	\tag{33}
\end{equation}
where \( \Gamma \) is the particle creation rate, and \( H \) is the Hubble parameter. This equation expresses the conservation of total energy in an expanding universe with an additional source term due to particle production.

To isolate the temperature evolution, we make use of the thermodynamic identity~\cite{Lima:1992np}, calculated as follows:
\begin{equation}
	\frac{\partial \rho}{\partial n} = \frac{h}{n} - \frac{T}{n} \frac{\partial p}{\partial T}, \label{eq:thermo-identity}
	\tag{34}
\end{equation}
where \( h = \rho + p \) is the enthalpy density. Substituting this into Equation~\eqref{eq:energy-balance} and solving for \( \dot{T}/T \), one obtains the following:
\begin{equation}
	\frac{\dot{T}}{T} = c_s^2 \frac{\dot{n}}{n} = c_s^2 (\Gamma - 3H), \label{eq:T-evolution}
	\tag{35}
\end{equation}
where the speed of sound squared is defined as \( c_s^2 = \partial p / \partial \rho \). This equation shows that the temperature evolution is directly influenced by the competition between particle creation (\( \Gamma \)) and dilution due to cosmic expansion (\( 3H \)).

\textls[-30]{In the particular case where the matter fluid obeys a barotropic equation of state of the following form}:
\begin{equation}
	p = (\gamma - 1) \rho, \qquad 1 \leq \gamma \leq 2, \label{eq:barotropic}
	\tag{36}
\end{equation}
the sound speed is constant: \( c_s^2 = \gamma - 1 \). The temperature then scales with the number density as follows:
\begin{equation}
	T = T_0 n^{\gamma - 1}, \label{eq:T-n-scaling}
	\tag{37}
\end{equation}
where \( T_0 \) is an integration constant. This result reflects the standard adiabatic behavior of ideal fluids in cosmology, modified here by the presence of gravitationally-induced \mbox{matter creation.}

These results indicate that the nonminimal coupling between curvature and matter not only leads to a dynamical creation of particles but also induces modifications in the thermal history of the universe. The deviation from standard temperature-redshift scaling may offer a potential observational window into such theories, particularly in the context of the cosmic microwave background and structure formation~\cite{Harko:2014pqa,Lima:2014hia}.

\subsection{Validity of the Second Law of Thermodynamics in Cosmology}

A fundamental requirement of any physically viable cosmological model is consistency with the second law of thermodynamics. This law asserts that the total entropy of an isolated system must be a non-decreasing function of time. In the context of an expanding universe taken as an open system, this principle must be generalized and \mbox{accordingly verified.}

\subsubsection{Thermodynamic Criteria}

In this context, to analyze this condition and establish specific thermodynamic criteria, one can define the total entropy \( S \), as measured by a co-moving observer, as the sum of two distinct contributions:
\begin{enumerate}
	\item The entropy associated with the apparent horizon, \( S_{\text{ah}} \);
	\item The entropy of the matter (and radiation) content inside the horizon, \( S_m \).
\end{enumerate}

Thus, the total entropy becomes the following:
\vspace{-6pt}
\begin{equation}
	S = S_{\text{ah}} + S_m. \label{eq:Stotal}
	\tag{38}
\end{equation}

In spatially flat FLRW cosmologies, the radius of the apparent horizon coincides with the Hubble horizon; therefore, it is given by \( r_{\text{ah}} = H^{-1} \), where \( H \) is the Hubble parameter. For a dust-dominated universe (i.e., a universe dominated by pressureless matter), the entropy can be expressed as follows:
\begin{equation}
	S = \frac{\pi}{H^2} + \frac{4\pi}{3H^3} n(t), \label{eq:Smatterhorizon}
	\tag{39}
\end{equation}
where \( n(t) \) is the number density of particles at time \( t \). The first term corresponds to the entropy of the apparent horizon, which is proportional to its area, while the second arises from the entropy of matter inside that horizon.

Therefore, to ensure consistency with the second law, the entropy must satisfy the following thermodynamic criteria:
\begin{equation}
S'(a) \geq 0, 
\label{eq:Sprime}
\tag{40}
\end{equation}
\begin{equation}
S''(a) \leq 0, 
\label{eq:Sdoubleprime}
\tag{41}
\end{equation}
where prime denotes derivation with respect to the scale factor \( a \).

\textcolor{black}{In cosmology, when considering the entropy $S(a)$ as a function of the scale factor $a$, the second law of thermodynamics requires that the entropy does not decrease, $S'(a) \ge 0$. Beyond this, thermodynamic stability imposes an additional condition on the second derivative, $S''(a) \le 0$, which corresponds to the concavity of the entropy function. This concavity condition ensures that the system evolves towards a state of maximum entropy: small fluctuations away from equilibrium decrease rather than increase the entropy, guaranteeing that the equilibrium configuration is stable. In a cosmological context, these conditions are naturally satisfied in a de Sitter phase, where the spacetime geometry is maximally symmetric, particle production ceases, and the entropy approaches its maximum value. Thus, $S'(a) \ge 0$ enforces the non-decreasing behavior required by the second law, while $S''(a) \le 0$ ensures that the entropy function is concave, confirming the stability and maximization of the entropy at equilibrium.}

\subsubsection{de Sitter Limit and Entropy Constraints}

A particularly illustrative case is when the universe evolves towards a de Sitter state, characterized by a constant Hubble parameter \( H = H_\star \). In this final phase, the scale factor grows exponentially, and the entropy behavior simplifies considerably. Assuming \( H = H_\star \), and using Equation~\eqref{eq:Smatterhorizon}, one finds that the first derivative of entropy is calculated \linebreak as follows:
\begin{equation}
	S'(a) = \frac{4\pi}{3 H_\star^4} \; \frac{n(a)}{a} \left[ \Gamma(a) - 3H_\star \right] \geq 0, \label{eq:SprimeG}
	\tag{42}
\end{equation}
which shows that particle creation enhances entropy. For the particular case \( \Gamma = 3H_\star \), we find that \( S' = 0 \), and hence \( S = \text{const} \), implying that the cosmological evolutions correspond to an isentropic process. This result is consistent with a pure de Sitter evolution, where the spacetime geometry is maximally symmetric and particle creation ceases. In addition, the second derivative, imposing the concavity condition, yields the following:
\begin{equation}
	S''(a) = \frac{4 \pi n(a)}{3 a^2 H_\star^5}
	\left\{
	\Gamma^2(a) + H_\star \left[ a \Gamma'(a) + 12 H_\star \right]
	- 7 H_\star \Gamma(a)
	\right\} \leq 0. \label{eq:SdoubleprimeG}
	\tag{43}
\end{equation}

The above conditions impose specific constraints on the functional form of the particle creation rate \( \Gamma(a) \) and its derivative \( \Gamma'(a) \), namely the following:
\begin{equation}
	\Gamma(a) \geq 3 H_\star, 
	\label{eq:Gammabound1} 
	\tag{44}
	\end{equation}
\begin{equation}	
	\Gamma'(a) \leq \frac{1}{a} \left[ 7 \Gamma(a) - \frac{\Gamma^2(a)}{H_\star} - 12 H_\star \right]. \label{eq:Gammabound2}
	\tag{45}
\end{equation}

Therefore, these constraints provide not only thermodynamic consistency but also dynamical bounds on particle production models in the context of modified gravity theories with nonminimal curvature--matter couplings. 

In fact, the validity of the second law in expanding universe models is a natural outcome of the deep connection between gravitation and thermodynamics, as emphasized in works on gravitational entropy and horizon thermodynamics~\cite{Padmanabhan:2009vy, Jacobson:1995ab}. In such interpretations, the apparent horizon plays the role of a causal boundary and acts analogously to the event horizon of black holes, carrying entropy proportional to its surface area. The consistency of entropy growth in cosmological spacetimes further supports the thermodynamic interpretation of gravity~\cite{Easther:1999gk}.

\subsection{Final Considerations}

The inclusion of matter creation mechanisms within cosmological models opens the door to novel and compelling scenarios for the evolution of the universe. In these frameworks, the universe may originate from an initial vacuum-like state, wherein matter and entropy are not present at the outset but are gradually generated through irreversible thermodynamic processes. This continuous macroscopic creation of matter effectively acts as a source of negative pressure, which can drive phases of accelerated cosmic expansion, thereby providing an alternative explanation to the standard dark energy paradigm. Importantly, these matter creation processes are deeply rooted in the principles of thermodynamics, offering a consistent mechanism for entropy production that respects the second law of thermodynamics. Furthermore, this approach is naturally framed within the context of non-equilibrium statistical mechanics, where the universe is treated as an open system.

\section{The Boltzmann Equation-Based Approach to Particle Creation} \label{secIV}

An alternative method for the investigation of the particle creation processes in cosmology, and generally in physics, is based on the extensive use of the methods of statistical physics and of the kinetic theory, using as a theoretical tool the Boltzmann equation. This approach was pioneered in \cite{Trig} and later extended and discussed in \cite{Lima, Bert,Lima1}.  In the following we will briefly review the physical foundations and the applications of the Boltzmann equation for the study of cosmological particle creation.

\subsection{The Boltzmann Equation}

The Boltzmann equation is a powerful mathematical method for the description of the physical details of the out-of-equilibrium processes \cite{Ver, Zhang}. It describes the evolution of the particle distribution as a result of the system evolution, including the effects of the momentum exchanges due to microscopic interactions. The Boltzmann equation relates the macroscopic and the microscopic evolution of the statistical distributions of the particles, and thus it can be used for the study of various physical situations. 

The Boltzmann equation gives the evolution in the phase-space with coordinates $\left(x^\mu,p^\mu\right)$ of the distribution function $f=f\left(x^\mu,p^\mu\right)$ of a thermodynamic system. It can be formulated in the following general form:
\begin{equation}
 \mathbf{L}[f]=\mathbf{C}[f],
 \tag{46}
\end{equation}  
where $\mathbf{L}$ and $\mathbf{C}$ are the Liouville and the collision operators, respectively.  
The Liouville operator describes the variation of the distribution function due to the motion of the particles, while the collision operator describes the effects of the microscopic processes on the evolution of $f=f\left(x^\mu,p^\mu\right)$. 

In the framework of a full general relativistic approach, the Liouville operator describes the evolution of the distribution function $f=f\left(x^\mu,p^\mu\right)$ along the geodesic line parameterized with the help of the affine parameter $\lambda$.  By using the definition of the momentum four-vector $p^\mu$, 
$ p^\mu=dx^\mu/d\lambda$, with the property $p^\mu p_\mu=m^2$, where $m$ is the mass of the particle, as well as the geodesic equation, written as follows:
\begin{equation}
 \frac{dp^\mu}{d\lambda}+\Gamma^\mu_{\nu \rho}p^\nu p^\rho=0,
 \tag{47}
\end{equation}
where $\Gamma^\mu_{\nu\rho}$ are the Christoffel symbols associated with the Levi-Civita connection, the Liouville operator is obtained as
\begin{equation}
 \mathbf{L}[f] \:\: \equiv \:\: \left.\frac{df}{d\lambda}\right|_{\rm geodesic \ line}
  = \frac{dx^\mu}{d\lambda}\frac{\partial f}{\partial x^\mu}+\frac{dp^\mu}{d\lambda}\frac{\partial f}{\partial p^\mu} 
  = p^\mu \frac{\partial f}{\partial x^\mu}-\Gamma^\mu_{\nu \rho}p^\nu p^\rho \frac{\partial f}{\partial p^\mu}.
   \tag{48}
\end{equation}

In the homogeneous and isotropic FLRW geometry, with the metric given by \mbox{$g_{\mu\nu}={\rm diag}\,(1,-a^2,-a^2,-a^2)$}, the distribution function $f$ is also spatially homogeneous and isotropic, so that $f=f(t,E)$, where $E$ is the energy of the particle. Then, the Liouville operator is obtained in the following form \cite{Zhang}:
\begin{equation}
 \mathbf{L}[f]=E\dot{f}-H|\vec{p}|^2\frac{\partial f}{\partial E}, 
  \tag{49}
\end{equation}
where $H=\dot{a}/a$ denotes the Hubble function, and $(\vec{p})^i\equiv ap^i$ is the $i$-th component of the physical momentum of the particle.

As for the collision operator $\mathbf{C}[f]$, giving the variation rate due to microscopic processes, it is defined according to \cite{Zhang}, results in the following:
\begin{equation}
 \mathbf{C}[f] \:\: \equiv \:\: \left.\frac{df}{d\lambda}\right|_{\rm microscopic \ process} \:\: = \:\: \left. E \: \frac{df}{dt}\right|_{\rm microscopic \ process}. 
  \tag{50}
\end{equation}

The evaluation of the distribution function is generally performed by assuming that the microscopic and collision processes can be described by quantum field theory. Furthermore, it is assumed that the microscopic interaction and collision processes can be studied in the Minkowski spacetime, with the gravitational effects included in the Liouville operator.

\subsection{The Boltzmann Equation in the Presence of Particle Creation}

The first model that includes gravitational particle production in the Boltzmann equation was proposed in \cite{Trig}, where the Boltzmann equation was modified to take the \mbox{following form}:
\begin{equation}
\mathbf{L}[f]\equiv p^\mu\frac{\partial f}{\partial x^\mu}-\Gamma^{\mu}_{\alpha \beta}p^\alpha p^\beta \frac{\partial f}{\partial p^\mu}=\mathbf{C}[f] + \mathbf{P}_{grav} \left(x^{\mu},p^{\mu}\right),
\tag{51}
\end{equation}
and the term $\mathbf{P}_{grav} \left(x^{\mu},p^{\mu}\right)$ is a non-collisional source term of gravitational/quantum origin, describing the gravitationally induced matter creation processes. However, in the following, we will present the approach developed in \cite{Lima}, in which the collisional term $\mathbf{C}[f]$ is neglected, since the particle creation term conceptually requires a different approach as compared to the collisional term.    

Hence, after neglecting the standard collisional term, the Boltzmann equation in the presence of particle creation becomes the following \cite{Lima}:
\begin{equation}\label{restricted}
 p^\mu\frac{\partial f}{\partial x^\mu}-\Gamma^{i}_{~\alpha \beta}p^\alpha p^\beta \frac{\partial f}{\partial p^i}=\mathbf{P}_{grav}. 
 \tag{52}
\end{equation} 

As for $\mathbf{P}_{grav}$, it is assumed that it is proportional to the Christoffel symbols, $\mathbf{P}_{grav}\propto \Gamma ^{\mu}_{\alpha \beta}$ (particle creation vanishes in flat spacetime), and $\mathbf{P}_{grav}\propto \Gamma /\Theta$ (condition suggested by the macroscopic thermodynamic approach), where $\Theta =3H$. 
Thus, $\mathbf{P}_{grav}$ is taken as proportional to the product of these two quantities, which gives the following \cite{Lima}:

\begin{equation}
\mathbf{P}_{grav}= -\lambda_b \frac{\Gamma}{\Theta}\Gamma^{\sigma}_{\alpha \beta}\bar{p}^\alpha \bar{ p}^\beta \frac{\partial f}{\partial \bar{p}^\sigma}, 
\tag{53}
\end{equation}
where $\lambda_b$ is a constant, and an overbar denotes the physical momentum $\left(\bar{p}^\mu\right)$, which is related to the co-moving momentum $p^\mu$ by the relation $\bar{p}^\mu =ap^\mu$ \cite{Lima, Zhang}. \textcolor{black}{This definition of the physical momentum, used in the context of the studies of the general relativistic Boltzmann equation, is different from the definition of the ``physical momentum'' as used in cosmology, in which the physical momentum $\tilde{p}$ of a particle is related to its co-moving momentum $p$ by the relation $\tilde{p}=p/a$.   
}

By adopting as the basic variables the physical momenta, with the overbars dropped, the generalized Boltzmann equation in the presence of matter creation is given by the following \cite{Lima}:
\begin{equation}\label{boltzmann}
 \frac{\partial f}{\partial t}=H\left(1-\frac{\Gamma}{\Theta} \right)p\frac{\partial f}{\partial p}.
 \tag{54}
\end{equation}

\subsection{Macroscopic Quantities}

Once the distribution function is known, the macroscopic thermodynamic quantities can be obtained according to the following definitions \cite{Stew,Bern, Cerc}:  

\begin{adjustwidth}{-\extralength}{0cm}
\begin{equation}\label{tmunu}
N^{\mu}=nu^\mu=\int{fp^\mu\frac{d^3p}{p^0}}, \qquad
S^\mu=su^\mu=-\int{\left(f \ln f - f\right)p^\mu\frac{d^3p}{p^0}},\qquad
T^{\mu \nu}=\int{fp^\mu p^\nu \frac{d^3p}{p^0}}.
\tag{55}
\end{equation}
\end{adjustwidth}

In the FLRW geometry,  the only nonvanishing components of particle flux $N^{\mu}$ and of the entropy flux  $S^{\mu}$ are  obtained as follows:
\begin{equation}
N^{0}=n=\int{f{d^3p}},\qquad
S^{0}=s=-\int{(f \ln f - f){d^3p}}. 
\tag{56}
\end{equation}

\textls[-25]{The energy momentum is diagonal, with the energy density defined as $\rho=u_{\mu}u_{\nu}T^{\mu\nu} \equiv T^{00}$.}

With the help of the above definitions, the divergences of the thermodynamic quantities can be easily obtained. For the divergence of the particle flux we find the \mbox{following \cite{Lima}}:
\begin{equation}
\begin{array}{ll}
 \nabla_\mu N^\mu \equiv \frac{1}{a^3}\frac{\partial}{\partial x^\mu}\left( a^3 \int{fp^\mu\frac{d^3p}{p^0}} \right)\\
 	{~} \\
 ~~~~~~~~~~~= \frac{1}{a^3}\frac{\partial}{\partial t}(a^3\int{fd^3p}) 
=3Hn+H\left(1-\frac{\Gamma}{\Theta}\right)\int{p\frac{\partial f}{\partial p}d^3p}.
\end{array}
\tag{57}
\end{equation}
where in the last step we have used Equation (\ref{boltzmann}). Since we have the following:
\begin{equation}
\int{p\frac{\partial f}{\partial p}d^3p}=4\pi \int{p^3\frac{\partial f}{\partial p}dp}=-3\int{fd^3p}=-3n,
\tag{58}
\end{equation}
we obtain the final result
\begin{equation}
 \nabla _\mu N^\mu=\dot{n}+3Hn=n\Gamma,
 \tag{59}
\end{equation}
which is precisely Equation~(\ref{n}), which was introduced in a phenomenological way. 

For the entropy flux we find the following:
\begin{equation}
\nabla _\mu S^\mu =-\frac{1}{a^3}\frac{\partial }{\partial t}\left(a^3 \int{(f \ln f - f)d^3p} \right), 
\tag{60}
\end{equation}
and
\begin{equation}
\nabla _\mu S^\mu=3sH-\int{\frac{\partial f} {\partial t}\ln f d^3p} 
=3sH-H\left( 1-\frac{\Gamma}{\Theta}\right)\int{p\frac{\partial f}{\partial p}\ln f d^3p}.
\tag{61}
\end{equation}

Since we have the following:
\begin{equation}
 \int{p\frac{\partial f}{\partial p}\ln f d^3p}=4\pi \int{p^3\frac{\partial f}{\partial p}\ln f dp} 
=-3\int{(f \ln f - f)d^3p}, 
\tag{62}
\end{equation}
 we obtain
\begin{equation}
\nabla _\mu S^\mu =-\Gamma\int{\left(f\ln f- f \right)d^3p}=s\Gamma
\tag{63}
\end{equation}

For the calculation of the divergence of the energy--momentum tensor one must assume, similarly to the macroscopic case, the existence of a non-collisional extra pressure term, which takes the form  $\Delta T^{i}_j  = - p_c\delta^{i}_j$, or equivalently, ${\Delta  T^{\mu \nu}}= - p_c h^{\mu \nu}$,  where $p_c$ is the creation pressure, introduced phenomenologically at this moment, and $h^{\mu \nu}=g^{\mu \nu}-u^\mu u^\nu$ is the projection operator. Hence we can write the following:
\begin{equation}
u_\mu \nabla_\nu T^{\mu \nu}\equiv u_\mu \left[ \frac{1}{a^3}\frac{\partial }{\partial x^\nu}(a^3T^{\mu \nu})+\Gamma^{\mu}_{\alpha \beta}T^{\alpha \beta}\right],
\tag{64}
\end{equation}
with $T^{\mu \nu} = T^{\mu \nu}_{coll}  + \Delta {\tilde T^{\mu \nu}}$, with $T^{\mu \nu}_{coll}$  given by  Equation~(\ref{tmunu}). After some simple calculations (\mbox{see \cite{Lima}} for the full details), we have the following:
\begin{equation}
u_\mu \nabla _\nu T^{\mu \nu}= 3Hp_c + (\rho+P)\Gamma.
\tag{65}
\end{equation}

Hence, the total matter energy momentum-tensor is divergenceless, if  and only if the creation pressure $p_c$ is given by the following:
\begin{equation}
p_c=-(\rho + P)\frac{\Gamma}{\Theta}.
\tag{66}
\end{equation}

\textls[-15]{In this way we have recovered the macroscopic Expression (\ref{pc}) for the creation pressure.}

\subsection{Temperature Evolution}

To obtain the temperature distribution of the cosmic fluid in the presence of matter creation we assume that the distribution function is given by the equilibrium form, written as follows \cite{Lima}:  
\begin{equation}
f=e^{\alpha(t)-\beta(t) E},
\tag{67}
\end{equation} 
where $\alpha(t)$ is a scalar function, and $\beta(t)=1/T(t)$ is the inverse of the temperature. Hence, from the Boltzmann Equation (\ref{boltzmann}) we find the following \cite{Lima}: 
\begin{equation}\label{65}
 \dot \alpha -\dot \beta E+ \beta H\left( 1-\frac{\Gamma}{\Theta} \right)\frac{p^2}{E}=0.
 \tag{68}
\end{equation}

For $\dot \alpha = {\dot \beta}=0$, the above equation has the trivial solution $\Gamma = \Theta = 3H$, describing an exponentially (de Sitter type) expanding universe, with $\dot \rho = {\dot n} = 0$. Hence, in this case the matter creation rate exactly compensates for the decrease of the matter density due to the expansion of the universe.

Two limiting solutions of Equation~(\ref{65}) can also be obtained, corresponding to the ultrarelativistic limit  ($m \ll T, E \simeq p$), and to the nonrelativistic limit ($m\gg T$, $E \simeq m+\frac{p^2}{2m}$), respectively. For the case  $m \ll T$, Equation~(\ref{65}) takes the following form \cite{Lima}:
\begin{equation}
 \frac{\dot \alpha}{\dot \beta}=E\left[1-\left( 1-\frac{\Gamma}{\Theta}\right)\frac{\dot a}{a}\frac{\beta }{\dot \beta} \right],
 \tag{69}
\end{equation}
which for $\dot \alpha =0$ has the solution
\begin{equation}
 \frac{\dot T}{T}=-\frac{\dot a}{a}+\frac{\Gamma}{3} \Leftrightarrow \frac{1}{aT}\frac{d (aT)}{dt} = \frac{\Gamma}{3},
 \tag{70}
\end{equation}
and
\begin{equation}
T=T_0 \left( \frac{a_0}{a} \right) e^{{\frac{1}{3}\int^{t_o}_{t}{\Gamma(x)}{dx}}}.
\tag{71}
\end{equation}

For the case  $m \gg T$, Equation~(\ref{65}) takes the following form:
\begin{equation}
 \frac{\dot \alpha}{\dot \beta}-m=\frac{p^2}{m}\left[ \frac{1}{2} - H \frac{\beta }{\dot \beta}\left( 1-\frac{\Gamma}{\Theta} \right) \right],
 \tag{72}
\end{equation}
with the solution  $\alpha - m \beta = \textrm{constant}$, and
\begin{equation}
 \frac{\dot T}{T}=-2\left( \frac{\dot a}{a} - \frac{\Gamma}{3}\right).
 \tag{73}
\end{equation}

For the discussion of the cosmological implications of these results see \cite{Lima}. 

\subsection{Alternative Approaches: Boltzmann Equation in the Presence of an Extra Force}

In modified theories with curvature--matter couplings the motion of massive particles is non-geodesic, as described by Equation~\eqref{eqmot}, this modification of the equation of motion was used in \cite{Bert} to suggest an alternative formulation of the Boltzmann equation, which is given by the following:
\begin{equation}
p^{\mu}\frac{\partial f}{\partial x^{\mu}}-\left(\Gamma^{\alpha}_{\mu\nu}p^{\mu}p^{\nu}-m^2F^{\alpha}\right)\frac{\partial f}{\partial p^{\alpha}}=\left(\frac{\partial f}{\partial \tau^*}\right)_{coll.},
\tag{74}
\end{equation} 
\textls[-20]{where $F^\alpha$  is the extra force generated by the curvature--matter coupling, defined in \mbox{Equation~(\ref{extra})}, and $\tau^*=\tau/m$ is the affine parameter. This approach leads to the modification of the Liouville operator in the presence of geometry matter coupling, so that we obtian the following \cite{Bert}}:
\begin{equation}
\mathbf{L}[f] =\left[p^{\mu}\frac{\partial }{\partial x^{\mu}}-\left(\Gamma^{\alpha}_{\mu\nu}p^{\mu}p^{\nu}-m^2F^{\alpha}\right)\frac{\partial }{\partial p^{\alpha}}\right] f. 
\tag{75}
\end{equation}

As for the collision term, the standard expression is adopted, so that we obtain the following:
\begin{equation}
\left(\frac{\partial f}{\partial t}\right)_{coll.}=\int{\left(f'_2f'-f_2f\right)\Phi \,\sigma \sqrt{-g}\frac{ d^3p_2}{(p_{2})_{0}}d\Omega},
\tag{76}
\end{equation}
where $\sigma$ is the differential cross-section,  $\Phi=\sqrt{\left[(p_2^{\mu}p_{2\;\mu})^2-m^4\right]}$ is the invariant flux, and  $d\Omega$ is the infinitesimal solid angle for the binary collisions, respectively. Moreover, we have denoted $f_2=f\left(x, p_2^\mu\right)$ and $f'_2=\left.\left(\partial f\left(x,p^\mu\right)/\partial p^\mu \right)\right|_{p^\mu\rightarrow p_2^\mu}$, respectively. 

A cosmological  application is also considered for the case of the FLRW metric, which gives for the Boltzmann equation the following form:
\begin{equation}
p^{0}\frac{\partial f}{\partial x^{0}}-2Hp^0p^i\frac{\partial f}{\partial p^{i}}=\int\left(f'_2f'-f_2f\right)\Phi \sigma d\Omega\sqrt{-g}\frac{d^3p_2}{(p_{2})_{0}},
\tag{77}
\end{equation}
which is identical to the Boltzmann equation for the FLRW metric in GR. Hence, in this approach the effects of the curvature--matter coupling appear due to the modifications of the Friedmann equations only. 

The extensive use of the Boltzmann equation can lead to a deeper understanding of the complex statistical and thermodynamical processes related to modified gravity theories in the presence of curvature--matter coupling and matter creation. However, as compared to the standard thermodynamic approach, no unique description, based on the Boltzmann equation, of the particle creation in modified gravity theories does exist. In particular, the role played by the collision term in the equation still deserves further investigation. Moreover, exploring the relation between the thermodynamic approach and the statistical physics one may lead to a better understanding of the complex interaction of matter \mbox{with gravity.    } 

For a discussion of Boltzmann's $H$-theorem in the framework of modified gravity theories with nonminimal curvature--matter couplings, see \cite{Avelino}. The sufficient condition for the violation of Boltzmann's $H$-theorem was found, together with the equation for the evolution of Boltzmann's $H$ function in terms of the nonminimal coupling between matter and geometry. This equation is valid for a collisionless gas in the FLRW universe. The relation between the high entropy of the universe and the weakness of gravity in modified theories of gravity was also explored.

\section{Quantum Aspects of Particle Creation} \label{secV}

\textls[-15]{The description of the interaction of the classical gravitational fields with quantized matter fields of spin zero and one-half represents a major topic of interest in theoretical physics. One of the major discoveries arising from these studies is the fundamental result that the expansion of the universe can create particles out of the vacuum. This phenomenon was initially explored by Schrödinger \cite{Schrodinger}, who first pointed out the possibility that a time-dependent gravitational background could lead to spontaneous particle production. This pioneering idea was rigorously developed and formalized by Parker~\cite{Parker:1968mv,Parker:1969au,Parkerbook}, who demonstrated that quantum fields propagating in curved spacetimes with dynamic metrics do not admit a unique vacuum state. Consequently, the spacetime geometry can induce transitions from vacuum to multi-particle states, effectively producing particles out of \mbox{the vacuum. }}

This process is inherently linked to the nonstationarity of the gravitational field and has profound implications for early universe cosmology, including mechanisms for reheating after inflation and the generation of primordial perturbations. These foundational works laid the groundwork for subsequent studies of particle creation in a variety of cosmological and astrophysical settings. In the present section, we briefly review the fundamental concepts and results in this field, pointing out the basic features of the quantum mechanical creation of particles in time-varying gravitational fields. We begin our analyses with a discussion of the Klein--Gordon equation in modified gravity theories with nonminimal curvature--matter couplings, and then we proceed to the presentation of the quantum perspective on the particle creation by the gravitational fields in an expanding universe.

\subsection{The Klein--Gordon Equation in the Presence of Particle Creation}

An important theoretical question is to find an answer to the question of the possible effects of the gravitational field in the presence of matter creation on the microscopic evolution and dynamics of elementary particles. To investigate the implications of matter creation on microscopic physics, we consider the case of a scalar field $\varphi$, whose physical properties are described by the Klein--Gordon equation. We assume that the scalar field mediates microscopic interactions, like, for example, the Higgs field, which generates the masses of the elementary particles \cite{Wang}. From a physical point of view, we cannot rule out the possibility of an interaction between the force of gravitation and the microscopic scalar field. Moreover, the possibility of the nonminimal coupling between the scalar field and gravity cannot be neglected a priori. Hence, in the following, we consider the effects of modified gravity in the presence of a nonminimal curvature--matter coupling to \mbox{the scalar field.}

\textls[-15]{Several theoretical approaches exist for the inclusion of a nonminimal curvature--matter coupling in the Klein--Gordon equation. In the simplest case of the massless particles, the Klein--Gordon equation is invariant under the group of conformal transformations \cite{Wald}. The massive Klein--Gordon equation can also be transformed into a conformally invariant equation through the addition of the term $\xi R\varphi$, with $\xi=1/6$ \cite{Wald}. An extra term of the same form is obtained in quantum field theory by introducing a counter-term in the Lagrangian, which renormalizes in curved spacetime theories with interacting scalar fields \cite{Bunch}. }

In the presence of curvature-scalar field coupling, the generalized Klein--Gordon equation can be derived by using a variational approach. The Lagrangian of the theory can be constructed by including a nonminimal quadratic coupling of the scalar field to the Ricci scalar. Thus, the action $S_\varphi$ of the scalar field nonminimally coupled to gravity is given in a covariant form by the following equation \cite{Amendola, Rossi}:
\vspace{-6pt}
\begin{equation}
S_\varphi=-\frac{1}{2}\int{\left(\nabla _\mu \varphi \nabla ^\mu \varphi +\xi R\varphi ^2+\frac{m_0^2}{2}\varphi^2\right)\sqrt{-g}d^4x},
\tag{78}
\end{equation}
where  $\varphi$ denotes the scalar field, $\Box = \nabla _\mu \nabla ^\mu$ is the d'Alembert operator, $R$ is the Ricci scalar, $m_0$ is the mass of the scalar field particle, and $\xi$ is a dimensionless coupling \mbox{constant, respectively. }

Therefore, in the presence of a nonminimal coupling between the scalar field and geometry, the Klein--Gordon equation is obtained in the following form \cite{Amendola, Rossi}:
\begin{equation}\label{72}
    \left(\Box +m_0^2+\xi R\right)\varphi=0,
    \tag{79}
\end{equation}
where we have adopted the $(+,-,-,-)$ metric signature. As one can see from the above equation, the nonminimal curvature-scalar field coupling modifies the Klein--Gordon equation in a significant way. The modifications may have important implications on the dynamics and cosmological evolution of the elementary particles, and they could provide some observational/experimental signatures that may help test the effects of modified gravity.  
For $\xi=0$, Equation \eqref{72} reduces to the Klein--Gordon equation with minimal gravity-scalar field coupling, $ \left(\Box +m_0^2\right)\varphi=0$.

By taking the trace of Equation \eqref{FE} we obtain the following relation:
\begin{equation}\label{73}
   \Psi \left(R,L_m\right) R+3\Box \Psi \left(R,L_m\right)=\left[1+\lambda f_2(R)\right]T^{(m)},
   \tag{80}
\end{equation}
where $\Psi \left(R,L_m\right)=F_1(R)+2\lambda F_2(R){\cal L}_m$, and $T^{(m)}=g^{\mu \nu}T_{\mu \nu }^{(m)}$. 

Let us assume now that a solution $R=R\left(L_m,T^{(m)}\right)$ of Equation~(\ref{73}) is known. Substituting this solution into Equation~(\ref{72}) gives the modified Klein--Gordon equation in the framework of modified gravities with curvature--matter coupling as follows:
\begin{equation}
    \left(\Box +m_{\rm eff}^2\right)\varphi=0,
    \tag{81}
\end{equation}
where 
\begin{equation}
    m_{\rm eff}^2 = m_0^2+\xi R\left(L_m,T^{(m)}\right).
    \tag{82}
\end{equation}

Therefore, $m_{\rm eff}$ represents the effective mass of the scalar field in modified gravities with a curvature--matter coupling. The scalar field interacts not only with its own mass $m_0$, but also with the matter Lagrangian and the trace of the matter energy--momentum tensor, as shown by the additional term in the mass equation. The generalization of the Klein--Gordon equation leads to a more complicated scalar field evolution and can provide deeper insights for the understanding of cosmological and astrophysical phenomena, including inflation, late-time cosmic acceleration, dark energy, or the properties of compact general relativistic objects.

\subsection{Particle Creation: The Quantum Perspective}

Particle creation is also a specific quantum mechanical process and one of the main characteristics of quantum field theory in curved spacetime \cite{Birrell:1982ix, Schrodinger, Parker:1968mv,
Parker:1969au, Parkerbook}. For a recent review on the quantum aspects of particle creation, see \cite{Ford}. In the present subsection, for the sake of completeness, we will briefly present the foundations of the quantum field theory approach to matter creation. 

Quantum particle creation is usually described in the Heisenberg picture by assuming that the state vector in the expanding universe, corresponding to the free quantized field, does not contain particles at early stages of evolution. If one requires the state vector at late times to contain a non-zero number of particles, then the annihilation operators of the early-time quantized field must evolve at late times into the linear combinations of annihilation and creation operators of the field \cite{Parker1}.  

Thus, in order for quantum particle creation to take place, the Heisenberg equation of motion of the field must evolve in time, together with the quantized field $\phi$. The second requirement is that the late time annihilation operator $B_{\vec k}$ is a linear combination of the initial annihilation and creation operators $A_{\vec k}$ and $A_{-\vec k}{}^{\dagger}$, given by the following:
\begin{equation}
 B_{\vec k} = \alpha_{k}\, A_{\vec k} + \beta_{k}\, A_{-\vec k}{}^{\dagger}.  
 \tag{83}
 \end{equation}
 
 This Bogoliubov-type transformation describes, within the framework of the momentum conservation, the creation of particle-antiparticle pairs of total zero momentum from the vacuum \cite{Parker1}.  More general Bogoliubov transformations, involving a sum over different momenta, of the following form:
\begin{equation}
 B_{\vec k} =\sum_{\vec k'} \alpha_{k, k'}\, A_{\vec k'} + \beta_{k, k'}\, A_{-\vec k'}{}^{\dagger},
 \tag{84}
 \end{equation}
also appear in physical processes related to gravitationally induced particle creation.

\subsubsection{The Free Scalar Field  in the Expanding Universe}\label{sec-free}

Let us first consider the case of the free, minimally coupled quantized scalar field $\phi$ in the FLRW geometry with the following metric:
\begin{equation}\label{KG-1} 
ds^2 = dt^2 - a^2(t)\left(dx^2 + dy^2 + dz^2\right),
\tag{85}
\end{equation}
with coordinates $\left(t,x^i\right)=(t,x,y,z)$. The Klein--Gordon equation satisfied by the scalar field is as follows:
\begin{equation} 
(\Box + m_0^2)\phi = 0, \label{KG-2}
\tag{86}
\end{equation}
where $\Box= g^{\mu \nu} \nabla _\mu \nabla _\nu$, with $\nabla$ denoting the covariant derivative. By considering the metric in Equation (\ref{KG-1}), the Klein--Gordon equation becomes the following:
\begin{equation}\label{KG-3} 
\left[\frac{1}{a^3} \frac{\partial }{\partial t}\left(a^3\frac{\partial }{\partial t}\right) -\frac{1}{a^2}  \Delta + m_0^2 \right]\phi = 0, 
\tag{87}
\end{equation}
where we denoted the Laplacian operator as $\Delta =\frac{\partial ^2}{\partial x^2}+\frac{\partial ^2}{\partial y^2}+\frac{\partial ^2}{\partial z^2}$. 

In order to solve Equation~(\ref{KG-3}), we consider first the mathematical problem in the presence of periodic boundary conditions in a cube with
sides of length $L$, and volume $V = L^3$. After the physical quantities are obtained  we take the limit $L\rightarrow \infty$. In this case, we expand the field operator $\phi$ according to the following \cite{Parker1,Parkerbook}:
\begin{equation}\label{KG-4}
\phi = \sum _{\vec k}\left\{ A_{\vec k} f_{\vec k}(\vec x, t)
+ A^\dagger_{\vec k} f^*_{\vec k}(\vec x, t)\right\}, 
\tag{88}
\end{equation}
where
\begin{equation}\label{KG-5} 
f_{\vec k} = \frac{1}{\sqrt{2V a^3(t)}} \, e^{i\vec k\cdot\vec x} \, h_k(t),
\tag{89}
\end{equation}
where $n^i$ an integer, $k^i = 2\pi n^i/L$, $k = \vert\vec k\vert$. The quantities  $h_k(t)$ satisfy the second order ordinary differential equation, written as follows \cite{Parker1}:
\begin{equation}\label{KG-6}
 {d^2\over{dt^2}} h_k +  {k^2\over a^{2}} h_k(t) + m_0^2 h_k(t ) - {3\over 4} \left({\dot a \over a}\right)^2 h_k(t)
  - {3\over 2} {\ddot a \over a} h_k(t) = 0. 
  \tag{90}
\end{equation}

Moreover, the coefficients $f_{\vec k}$ satisfy the normalization conditions, written as follows:
\begin{equation}\label{KG-7} 
\left(f_{\vec k},f_{\vec k'}\right) = \delta_{\vec k,\vec k'}, 
\left(f_{\vec k}, f_{\vec k'}^*\right) = 0, 
\tag{91}
\end{equation}
where  on the left-hand-side of Equation (\ref{KG-7}) the brackets denote scalar products, which are conserved quantities. The scalar products must be computed through integration over a spatial volume $V$ at time $t$ in the FLRW geometry. 
 The creation and annihilation operators $A_{\vec k}$ satisfy the commutation relations, written as follows:
\begin{equation} 
\left[ A_{\vec k}, A_{\vec k'}\right] = 0, \qquad
  \left[A^\dagger_{\vec k}, A^\dagger_{\vec k'}\right] = 0 , 
  	\qquad
\left[ A_{\vec k}, A^\dagger_{\vec k'}\right] = \delta_{\vec k,\vec k'} . \label{2-6a}
\tag{92}
\end{equation}

Analytic solutions of Equation~(\ref{KG-6}) can be obtained for the cases of power and exponential
expansions. For the  case of the exponential expansion,  $a(t) =\exp (H_0t)$, where  $H_0$ is the Hubble constant, the solutions of Equation~(\ref{KG-6}) can be expressed in terms of  Hankel functions $H_\nu^{(n)}$, and are given by the following \cite{Parker1}:
\begin{equation}
f_{\vec{k}} = \frac{1}{2V a^3(t)}e^{\vec{k}\cdot \vec{x}}H^{(1)}_\nu(v),
\tag{93}
\end{equation}
where
\begin{equation}
v=\frac{k}{H_0} e^{-H_0t}, \qquad \nu=\sqrt{\frac{9}{4}-\frac{m_0^2}{H_0^2}}.
\tag{94}
\end{equation}

An approximate solution of Equation~(\ref{KG-6}) can be obtained in the adiabatic approximation \mbox{as follows \cite{Parker1, Parkerbook}}:
\begin{equation}\label{KG-8}
h_k(t) \sim \frac{1}{\sqrt{\omega _k(t)}}e^{\pm i\int^t{\omega _k\left(t'\right)dt'}}, 
\tag{95}
\end{equation}
with
\begin{equation}
\omega _k (t)=\left[\frac{k}{a^2(t)}+m_0^2\right]^{1/2}.
\tag{96}
\end{equation}

The two solutions present in Equation (\ref{KG-8}) are linearly independent, and thus the general solution of Equation~(\ref{KG-6}) is given \mbox{by the following \cite{Parker1}}:
\begin{equation}
h_k (t)\approx  \frac{\alpha _k}{\sqrt{\omega _k(t)}}e^{- i\int^t{\omega _k\left(t'\right)dt'}}+\frac{\beta _k }{\sqrt{\omega _k(t)}}e^{+ i\int^t{\omega _k\left(t'\right)dt'}}, 
\tag{97}
\end{equation} 
where $\alpha _k$ and $\beta_k$ are complex constants satisfying the relation
\begin{equation}
\left|\alpha _k\right|^2-\left|\beta _k\right|^2=1.
\tag{98}
\end{equation}

The expectation value of the number of particles created due to the expansion of the universe from the early-time vacuum state $|0>$ to the present time 
 is as follows \cite{Parkerbook}:
 \begin{equation}
\left< 0\left|B_{\;\vec{k}}^{+}B_{\vec{k}}\right|\right>=\left|\beta _k\right|^2, 
\tag{99}
 \end{equation}
 where 
 \begin{equation}
 B_{\vec{k}}=\alpha _k A_{\vec{k}}+\beta _k A_{\;-\vec{k}}^{+}.
 \tag{100}
 \end{equation}
 
Therefore, except in a few special cases, the expansion of the universe creates quanta of the minimally coupled scalar field. In order to conserve the 3-momentum, in the FLRW universes the particles are created in pairs \cite{Parker1}.

Finally, in the case of an exponentially expanding FLRW universe, one obtains the following relation \cite{Parker1}:
\begin{equation}
\left|\beta _k\right|^2\propto e^{3H_0t}. 
\tag{101}
\end{equation}

\subsubsection{The Parker--Toms Model}

As another particular quantum particle creation model we consider the Parker--Toms model \cite{Ford, PT1,Parkerbook}. This model is given by an exact solution solution 
 for the creation of massless, minimally coupled scalar particles. In order to obtain it one transforms the time coordinate $t$ to a new time coordinate $\tau$ according to the relation $dt = a^3 d\tau$. In the new time coordinate FLRW metric takes the following form:
 \begin{equation}
ds^2 = - a^6(\tau) d\tau^2 + a^2(\tau)\left(dx^2+dy^2+dz^2\right).
\tag{102}
 \end{equation}
 
 We look now for the solutions of the  minimally coupled, massless Klein--Gordon equation $\Box \varphi =0$. The solution of the equation can be written as follows:
 \begin{equation}
f_{\bf k}({\bf x},\tau)= \frac{e^{i{\bf k\cdot x}}}{\sqrt{(2\pi)^3}}\,\, \chi_k(\tau), 
\tag{103}
\end{equation}
which is independent on the scale factor. 
 The functions $\chi_k(\tau)$ satisfy the following equation:
\begin{equation}\label{KG-9}
\frac{d^2\chi_k(\tau)}{d\tau^2} +  k^2 \, a^4(\tau) \chi_{k}(\tau) =0.
\tag{104}
\end{equation}

Parker and Toms \cite{PT1,Parkerbook} proposed the following expression for the scale factor:
  \begin{equation}
 a^4(\tau) = \frac{1}{2} \left[1 + a_i^4 +(1-a_i^4) \tanh(\sigma \tau)\right],
 \tag{105}
 \end{equation}
where $\sigma$ is a constant. In this model, the universe expands from an initial value  $a_i < 1$ of the scale factor to a final value  $a_f =1$. With this scale factor Equation~(\ref{KG-9}) becomes \linebreak the following:
\begin{equation}
 {\frac{d^2\chi_k(\tau)}{d\tau^2}} + \frac{1}{2}  k^2   \left[1 + a_i^4 +\left(1-a_i^4\right) \tanh(\sigma \tau)\right] \chi_{k}(\tau) =0.
 \tag{106}
  \end{equation} 
  
The Bogoliubov coefficients can be found as follows:
  \begin{equation}
  | \beta_k|^2 = \frac{  \sinh^2[\pi k \left(1-a_i^2\right)/(2\sigma)]}{ \sinh(\pi  k/\sigma) \sinh(\pi k a_i^2/\sigma)}.
  \tag{107}
 \end{equation}
 
 In the high frequency limit, $k > k a_i^2 \geq \sigma$, we obtain the following:
  \begin{equation}
  \left| \beta_k\right|^2 \approx  e^{-2 \pi k a_i^2/\sigma},
  \label{eq:N-PT2}
  \tag{108}
 \end{equation}
 a result indicating that an exponential suppression of the creation rate of particles with energies larger than the inverse expansion time $\sigma$ does occur in this model. 
 
The total number $n$ and the energy density $u$ of the created particles can be obtained as follows:
  \begin{equation}
 n \approx \frac{\sigma^3}{8 \pi^5  a_i^6}, \qquad u \approx \frac{3 \sigma ^4}{16 \pi^6  a_i^8}.
 \tag{109}
 \end{equation}
 
 The mean energy $\left<E\right>$ of the created particles is as follows:
  \begin{equation}
\left<E\right> = \frac{u}{n} \approx \frac{3 \sigma}{2 \pi  a_i^2}.
\tag{110}
 \end{equation}
 
 Hence, the Parker--Toms model provides an exact solution for the creation of massless, minimally coupled scalar particles in a universe in bounded expansion. Moreover, this solution can also be used to describe graviton production in the expanding universe. Moreover, the concrete estimates for the particle creation rates and of their number and energy densities may open the possibility of the comparison with some astrophysical or cosmological observational data. Quantum matter creation processes may have played an important role in the creation of particles in the universe after the end of an inflationary era, during the reheating period, and they could also explain the mystery of dark matter \cite{Ford}. The creation of gravitons by quantum field theoretical processes could also have important astrophysical and cosmological implications, since the energy density of the newly created gravitons leads to an increase of the total energy density of the universe, and thus it may influence the cosmological expansion rate. Hence, particle creation processes could be directly constrained via observations of Type Ia supernovae, CMB, or even high-energy astrophysical phenomena.

\section{Conclusions}\label{secconclusion}

In this work, we examined the cosmological implications of modified theories of gravity that incorporate a linear coupling between the Ricci scalar curvature and the matter Lagrangian. A distinctive feature of these models is the non-conservation of the energy--momentum tensor, which leads to the possibility of particle creation induced by the gravitational field. Using the scalar--tensor representation of the theory, we derived the particle creation rate, the corresponding creation pressure, and the entropy production rate, thereby characterizing the irreversible energy transfer from the gravitational sector to the matter content of the universe.

\textcolor{black}{An important result of observational cosmology is represented by the fact that the universe entered, in its late phases of evolution, into an accelerated expansion state, which may end in a de Sitter-type exponential expansion. Therefore, realistic and viable cosmological models must also allow for the presence of an exponentially \mbox{expanding solution.}}

In the models with geometry--matter coupling the cosmic acceleration, as well as its de Sitter phase, is a natural result of the matter creation processes. This outcome follows from the generalized energy conservation Equation (\ref{eq:modified_energy}), which, for a pressureless fluid gives for the particle creation rate the following expression:
\begin{equation}
\Gamma =3H+\frac{\dot{\rho}}{\rho}.
\tag{111}
\end{equation}

The de Sitter expansion corresponds to $H=H_0={\rm constant}$, which allows to obtain the matter density as a function of the particle creation rate in the following form:
\begin{equation}
\rho(t)=\rho_0e^{\int{\Gamma (t)dt}-3H_0t},
\tag{112}
\end{equation}
where $\rho_0>0$ is an arbitrary integration constant. Hence, if matter creation takes place on large scales in the universe, a de Sitter type expansion can be obtained even in the presence of a nonzero matter density. If $\Gamma =3H_0={\rm constant}$, then $\rho (t)=\rho_0={\rm constant}$, that is we obtain a model of the constant density universe exponentially expanding in space.  For this case, the creation pressure becomes $p_c=-[(\rho+p)/3](\Gamma /H)=-\rho_0$.

\textcolor{black}{ Hence, our analysis below shows that the modified gravitational field equations in the presence of a geometry--matter coupling do admit de Sitter-type accelerating solutions at late times. This cosmic acceleration emerges naturally from the matter creation process, which induces a negative creation pressure. Therefore, in this framework, the observed late-time acceleration of the universe can be interpreted not as evidence for dark energy, but rather as a manifestation of gravitationally induced matter creation. This scenario offers a compelling alternative to the standard $\Lambda$CDM paradigm.
}

Furthermore, the thermodynamic consistency of the theory was thoroughly investigated in the context of irreversible processes associated with particle creation. We demonstrated that the total entropy of the universe, defined as the sum of the entropy of the apparent cosmological horizon and the entropy of the matter and radiation fields enclosed within it, exhibits a monotonically increasing behavior throughout cosmic evolution. As the universe approaches asymptotically the final de Sitter phase, the total entropy tends to a maximum, indicating that the system evolves toward thermodynamic equilibrium. This behavior provides strong support for the validity of the generalized second law of thermodynamics in models with curvature--matter coupling and particle production. Importantly, the rate of entropy production imposes significant and non-trivial constraints on the particle creation rate $\Gamma$ and on its functional dependence on the scale factor. These thermodynamic conditions must be satisfied in order to ensure not only the internal consistency of the theory but also its compatibility with a physically viable cosmological evolution.

One of the most intriguing and interesting physical characteristics of modified gravity theories with nonminimal curvature--matter couplings is the possibility of the presence of gravitational matter creation at both astrophysical and cosmological scales. This phenomenon requires a physical understanding and interpretation that goes beyond standard gravity, and it must be explored by using different approaches already developed in different fields of physics. In the present review we have briefly introduced and described three distinct descriptions of matter creation processes, represented by the classical thermodynamics of irreversible matter creation in open systems, the Boltzmann equation-based approach, and the creation of particles in an expanding universe. If the thermodynamic description of matter creation, initiated in \cite{Prigogine:1988jax}, gives a consistent, elegant, and systematic approach to the particle production in gravitational fields, it still retains a phenomenological/empirical character, with many important problems unanswered, like, for example, the nature of the created particles, their physical state, or the physical causes of these processes. On the other hand, the Boltzmann equation approach introduces a statistical perspective and a more physical description of the creation process. However, as expected, under certain assumptions, the Boltzmann equation approach leads to a complete equivalence with the open systems thermodynamical formulation \cite{Lima}. However, this equivalence does not impose any theoretical restrictions on the applications of the Boltzmann equation to the problem of matter creation, since the large spectrum of the applications of statistical physics to the problem of particle production could lead to major advances in the field. From a theoretical point of view, further studies are necessary to understand the role of the collision term in the Boltzmann equation in the presence of matter creation---can this term be safely neglected, since it may describe some weak physical effects, or does it have a more fundamental meaning, role, and interpretation?  

Of special interest in the framework of modified gravity theories with curvature--matter coupling are the quantum particle creation processes related to the expansion of the universe \cite{Ford,PT1,Parkerbook}, and generally to the presence of time-varying gravitational fields. These phenomena are a direct consequence of the quantization in an FLRW geometry of the basic evolution equations of particle physics, like the Klein--Gordon and the Dirac equations. In the presence of a gravitational field, the Klein--Gordon equation can be further generalized by adding in the action a term proportional to the Ricci scalar and coupled to the scalar field. This term generates in the Klein--Gordon equation an effective, time-dependent mass term. By using the more general representation of the Ricci scalar as obtained from the field equations of modified gravity theories with curvature--matter coupling, a generalization of the Klein--Gordon equation can be obtained in the presence of gravitational fields described by these theoretical approaches. This generalization is essentially related to the modification of the effective mass of the particle, which now also includes the curvature--matter coupling. It is certainly of interest to obtain solutions of this generalized Klein--Gordon equation and to explore their physical implications.

However, the most interesting aspect of the scalar field Klein--Gordon equation is related to its quantization in the FLRW geometric background, which directly leads to the physical process of particle production. Once this analysis is performed for the Klein--Gordon equation in the extended gravitational field theories, a direct relation may be established between the quantum properties of matter creation, described by the Bogoliubov coefficient $\beta _k$, and the macroscopic quantities characterizing the properties of the extended gravitational field. This approach could open a new perspective on the understanding of the relation between particle creation, thermodynamics, modified gravity, and quantum field theory.       

The curvature--matter coupling function $U(\psi)$, which characterizes and governs the nonminimal interaction between geometry and matter, may, in principle, originate from fundamental quantum field theoretical models of gravitation. These functions encapsulate the dynamical imprint of the underlying microphysical processes that could mediate the exchange of energy and momentum between spacetime curvature and the matter content of the universe. Their form could be dictated by symmetry principles, effective field theory arguments, or even by specific mechanisms in candidate theories of quantum gravity, such as string theory or loop quantum gravity. This theoretical possibility opens a compelling avenue for embedding the curvature--matter coupling framework into a more fundamental high-energy description of gravity. Moreover, it enables a rigorous confrontation of the predictions of these models with high-precision observational data from cosmology and astrophysics, thereby providing a pathway to test the viability and physical relevance of such extensions of GR in the current and future observational landscape.

To assess the physical viability of nonminimal curvature--matter coupling models, it is essential to establish observational criteria that can distinguish them from standard dark energy scenarios, including those that reproduce similar background expansion histories. Although such models may be degenerate with $\Lambda$CDM at the level of the Hubble parameter or the scale factor evolution, they can lead to distinct signatures in the evolution of cosmic structures. Key observational probes such as the large-scale distribution of galaxies, the anisotropies in the cosmic microwave background radiation, weak and strong gravitational lensing patterns, redshift-space distortions, and the linear and nonlinear growth rate of cosmological perturbations offer promising avenues for empirical discrimination. These observables are sensitive to the underlying gravitational dynamics and matter creation processes and therefore provide a powerful means to constrain or falsify modified gravity models with nonminimal curvature--matter coupling in the era of precision cosmology.

Finally, the forthcoming generation of astronomical surveys, such as \textit{EUCLID}, the \textit{Square Kilometre Array (SKA)}, the \textit{Dark Energy Survey (DES)}, and the \textit{Extended Baryon Oscillation Spectroscopic Survey (eBOSS)} as part of \textit{SDSS-III}, will provide high-precision cosmological data capable of mapping the expansion history, cosmic large-scale structure, and gravitational lensing signatures with unprecedented detail. These observational breakthroughs will enable rigorous tests of modified gravity models, including those with nonminimal curvature--matter couplings, and may offer the first concrete evidence for---or against---gravitationally induced matter creation. In this new era of precision cosmology, the synergy between theory and observation has the potential to uncover profound insights into the nature of cosmic acceleration, the elusive components of dark matter and dark energy, and the fundamental interactions that shape the evolution of the universe. With these tools at our disposal, we stand at the threshold of a deeper understanding of \mbox{gravity itself.}

\vspace{6pt} 
\authorcontributions{All the authors have substantially contributed to the present work. All authors have read and agreed to the published version of the manuscript.}
\funding{FSNL and MASP acknowledge support from the Funda\c{c}\~{a}o para a Ci\^{e}ncia e a Tecnologia (FCT) Research Grants UIDB/04434/{2020} (\url{https://doi.org/10.54499/UIDB/04434/2020}), UIDP/04434/2020 (\url{https://doi.org/10.54499/UIDP/04434/2020}) and PTDC/FIS-AST/0054/2021 (\url{https://doi.org/10.54499/PTDC/FIS-AST/0054/2021}).
MASP also acknowledges support from the FCT through the Fellowship UI/BD/154479/2022 (\url{https://doi.org/10.54499/UI/BD/154479/2022}). MASP also acknowledges financial support from the Spanish Grant PID2023-149560NB-C21, funded by MCIN/AEI/10.13039/501100011033.} 


\dataavailability{{No} new data were created or analyzed in this study.} 
\acknowledgments{FSNL acknowledges support from the FCT Scientific Employment Stimulus contract with reference CEECINST/00032/2018.}

\conflictsofinterest{The authors declare no conflicts of interest.} 






\begin{adjustwidth}{-\extralength}{0cm}

\reftitle{References}

\PublishersNote{}
\end{adjustwidth}
\end{document}